\documentclass[journal]{IEEEtran}
\usepackage{amsmath}
\usepackage{graphicx}
\usepackage{cite}
\usepackage{amsfonts}
\usepackage{booktabs}
\usepackage{multirow}
\usepackage{flushend}
\usepackage{mathrsfs}
\usepackage{hyperref}
\usepackage{cleveref}

\usepackage{longtable, booktabs}
\usepackage{supertabular}
\usepackage{lscape}
\usepackage{eurosym}
\usepackage{verbatim}
\usepackage{soul, color, xcolor}
\ifCLASSINFOpdf
\else
\fi

\begin{document}

\title{The ITU Vision and Framework for 6G: \\ Scenarios, Capabilities and Enablers}

\author{
Ruiqi Liu, Leyi Zhang, Ruyue Yu-Ngok Li and Marco Di Renzo  

\thanks{
R. Liu, L. Zhang and R. Li are with State Key Laboratory of Mobile Network and Mobile Multimedia Technology, Shenzhen, China and ZTE Corporation, Shenzhen, China (e-mail: richie.leo@zte.com.cn, leyi.zhang@zte.com.cn, li.ruyue@zte.com.cn).
M. Di Renzo is with Universit\'e Paris-Saclay, CNRS, CentraleSup\'elec, Laboratoire des Signaux et Syst\`emes, 3 Rue Joliot-Curie, 91192 Gif-sur-Yvette, France (e-mail: marco.di-renzo@universite-paris-saclay.fr). \\
Corresponding author: R. Li.
}
}

% make the title area
\maketitle
\begin{abstract}
With the standardization and commercialization completed at an unforeseen pace for 5th generation (5G) wireless networks, researchers, engineers and executives from the academia and industry have turned their attention to new candidate technologies that can support next generation wireless networks enabling more advanced capabilities in emerging scenarios. Explicitly, the 6th generation (6G) terrestrial wireless network aims to providing seamless connectivity not only to users but also to machine type devices for the next decade and beyond. This paper describes the progresses moving towards 6G, which is officially termed as ``international mobile telecommunications (IMT) for 2030 and beyond'' in the International Telecommunication Union Radiocommunication Sector (ITU-R). Specifically, the usage scenarios, their representative capabilities, the supporting  technologies and spectrum are discussed, and {research} opportunities and challenges are highlighted.
\end{abstract}

\IEEEpeerreviewmaketitle

\section{Introduction}
The history of human beings communicating through radio waves without a wire started in the 1890s, when physicists Alexander Popov and Guglielmo Marconi individually demonstrated the wireless telegraph. Generation through generation, the functions of communication systems have evolved from analog voice calls and instant text messages to Internet surfing and live streaming, even including more sophisticated functions such as localization. Indeed, 5th generation (5G) wireless networks have strengthened the traditional connection service and supported many vertical industries. Witnessing the profound impact brought to our lives by 5G makes us wonder, what will the 6th generation (6G) be?

With the rapid roll-out of 5G in major cities around the world, the daily lives of people as well as vertical industries have changed tremendously. Data is available to more people at a reduced cost, through more stable connections and higher speeds. Industries such as manufacturing, healthcare, and transportation also benefit from enhanced connectivity. While 5G technology offers many advantages, including increased capacity, reduced latency, enhanced reliability and higher throughput, there are several challenges associated with using 5G in industries, especially for those verticals that have requirements beyond traditional telecommunications. Integrating 5G technologies into different industrial systems can also be challenging, particularly for those supporting complex operations and legacy systems. In addition, it is challenging from both an economic and environmental perspective to operate 5G networks in a more energy efficient manner. While the 3rd generation partnership project (3GPP) has been working on 5G-Advanced to tackle some of these challenges,  some of them are still open to research. It is expected that 6G will become a game changer by providing more integrated and intrinsic features that can be adapted to different needs for the benefit of the society.

While 6G will become a reality after 2030, research projects and standards development organizations (SDOs) have already started conceptual work. In this context, the international telecommunication union radiocommunication sector (ITU-R) is the official administrator under the ITU tasked with managing spectral resources worldwide and specifying standards for international mobile telecommunications (IMT) systems. There has been some recent progress in ITU-R on {the 6G framework}, as well as on using new spectrum. To benefit readers, our objective is to summarize the official 6G recommendations and decisions, including the associated usage scenarios, capabilities, key technology enablers as well as the potential spectrum usage.
%Based on the latest research results and the ITU-R progress, a relatively clear view of the scenarios, requirements, capabilities, key technology enablers as well as the spectrum of the 6G can be formed, which will be shared in the following sections.

\section{ITU-R Efforts in Defining 6G}\label{Sec:ITUR}
\begin{figure*}[t]
\centering
\includegraphics[width=0.98\linewidth]{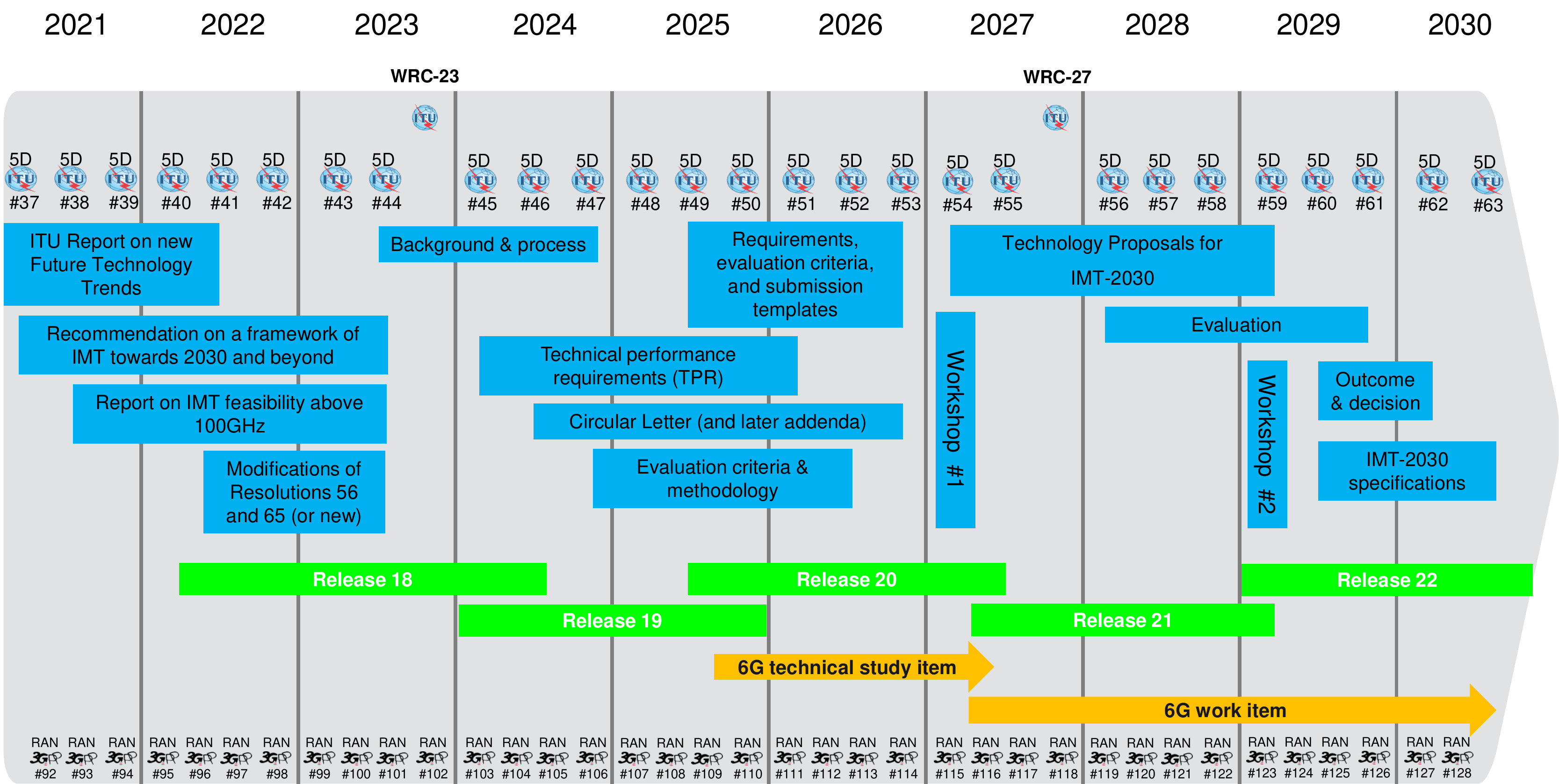}
\caption{Planned timeline for 6G studies and standardization in ITU-R WP5D and 3GPP RAN where the ITU-R programs, 3GPP releases and 3GPP phases are depicted in blue, green and orange, respectively. In general, 3GPP is the organization developing the detailed technical standards, while ITU-R is in charge of setting high-level goals for 6G and evaluating the technical standards.}
\label{fig:timeline}
\end{figure*}

{Working aside the 3GPP, whose task is mainly to develop technical specifications, the } ITU-R working party 5D (WP5D) is responsible for the development and evaluation of terrestrial IMT systems, with 6G currently being intensively studied. As depicted in Fig. \ref{fig:timeline}, to fully define a new generation of wireless networks, there are many chronological phases. Within this process, there are some key deliverables to be published, {including} the future technology trends (FTT) report, the recommendation on a framework of IMT-2030, and the report on technical performance requirements (TPR). Evaluation is also a critical step, starting with the TPR and the report on requirements, evaluation criteria and submission templates.

%There are three common types of deliverable published by WP5D, namely, the report, the recommendation and the resolution. Reports usually contain technical, operational or procedural statements and are treated as technical documents without strong enforcement. On the other hand, recommendations normally constitute a set of international technical standards on topics such as use of the spectrum / orbit resource, radio systems characteristics and performance, spectrum monitoring and emergency communication, and are treated with more weight by the other SDOs. Out of the three, resolutions are the most formal and need to be followed by all ITU members, which usually define high-level and non-technical aspects such as the official naming of different generations (5G being IMT-2020 and 6G being IMT-2030).

WP5D completed {and published} the FTT report in November 2022 \cite{ITU_FTT}. This report intends to provide views and study results on promising candidate technologies for 6G, and serves as a guideline for other SDOs. Some of the key technologies {discussed} in the FTT report {are} elaborated in Section \ref{sec:FTT}. 

Since early 2021, WP5D has been working on {a recommendation for framework of 6G}. %Later, its title was changed to a framework of IMT-2030 based on comments received during meetings, since a framework implies a more concrete meaning than vision. 
This recommendation has been completed during the 44th meeting of WP5D in June 2023 and approved by the Radiocommunication Assembly in November 2023 \cite{ITU_framework}. The key {aspects} of this recommendation include the usage scenarios and capabilities of 6G, which are detailed in Section \ref{Sec:usage_and_capabilities}. Beyond these two major parts, user application trends and potential interworking with previous IMT systems or other radio access systems are also {discussed} in this recommendation.  

As depicted in Fig. \ref{fig:timeline}, the conceptualization and study phase of 6G ends with the publication of the recommendation concerning the 6G framework, as well as by a subsequent and a more detailed technical phase, which is aimed at specifying minimal requirements for 6G systems and their evaluation. WP5D will release a series of reports, providing instructions for other SDOs to develop candidate 6G technologies, and submit their proposals to be evaluated by ITU-R. As a first step, WP5D will trigger work on the TPR, defining the minimal performance requirements, in correspondence with the  capability items in the framework recommendation. Since  preliminary studies on 6G have just started, the performance metrics provided in the framework recommendation are mostly a range of potential values to be considered, which will all be specified during the TPR phase. Furthermore, detailed deployment scenarios and network configurations (such as the system bandwidth, indoor or outdoor scenarios) supporting the capability metrics will also be specified. Simultaneously, WP5D will study the evaluation criteria and methodology, providing more details on how all the capability items are evaluated. Usually, several typical and representative deployment scenarios are considered to form a group of test cases, which can reflect the expected performance of the proposed technology. WP5D will also work on submission templates, {which are used by external SDOs (such as 3GPP) to propose technologies to be evaluated as potential 6G technologies}. The submission window for technology proposals for 6G will finally open in 2027, while the evaluation decisions will be made in 2029 the earliest. %Two workshops are planned to facilitate the consensus building process.

The 3GPP working group radio access network (RAN) is currently working on Release 19, which belongs to work of 5G-Advanced. This effort will continue at least in Release 20.  {Aiming at meeting the ITU-R requirements and passing the evaluation, 3GPP} study and self-evaluation phases of 6G will follow in Release 20, which is expected to start in mid-2025. The first release of 6G specifications will be Release 21 \cite{3GPP_timeline}, {which is targeted for the specification submission window of ITU-R WP5D}. 3GPP has agreed to freeze the first release of 6G no earlier than the first quarter 2029 \cite{3GPP_timeline}, and the timetable in Fig. \ref{fig:timeline} matches the previous history of having a new generation per decade.

\section{Usage Scenarios and Corresponding Capabilities}\label{Sec:usage_and_capabilities}
\subsection{Usage Scenarios}
\begin{figure}[t]
\centering
\includegraphics[width=0.85\linewidth]{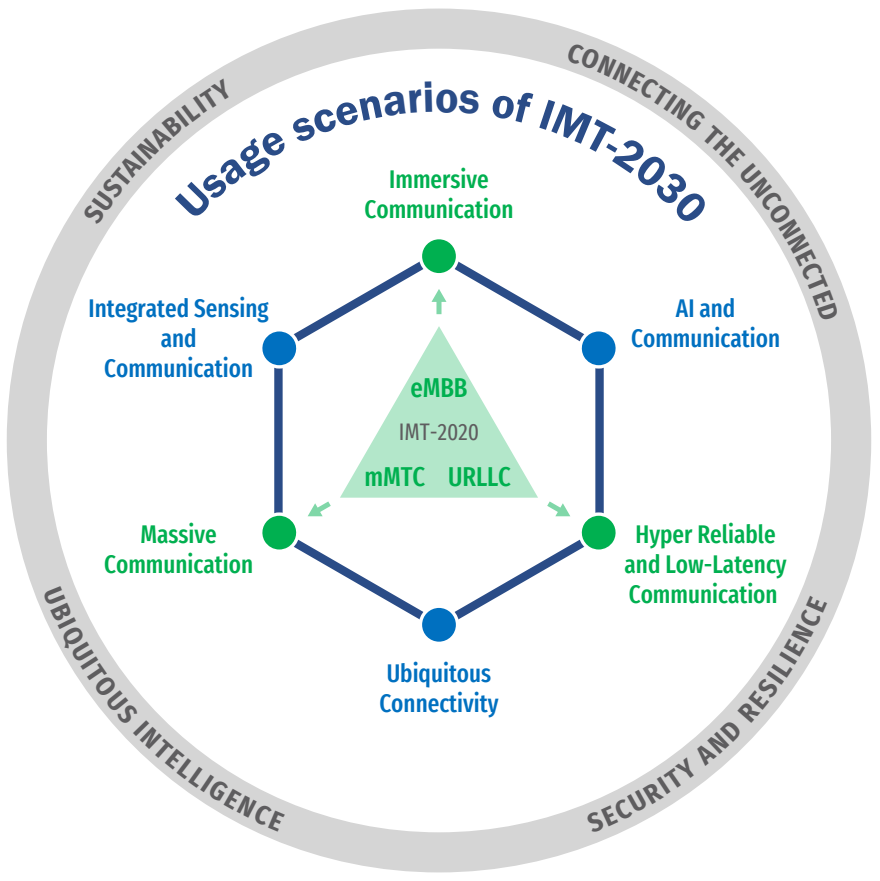}
\caption{A graphic illustration of the six usage scenarios and four overarching aspects of 6G approved by the ITU-R, the ``6G wheel''. (Source: \cite{ITU_framework})}
\label{fig:usage}
\end{figure}

%In the era of 6G, there are some new trends in user applications, which foster novel usage scenarios and demand new capabilities at the same time. 
Usage scenarios are a collection of use cases which have common aspects or similarities, and can be used as services. {In 5G}, there were three usage scenarios identified based on the user demands at that time, namely, enhanced mobile broadband (eMBB), massive machine-type communications (mMTC) {and} ultra-reliable and low latency communications (URLLC). In the time-frame of 6G, these three usage scenarios are expected to evolve into a more advanced version. As presented in Fig. \ref{fig:usage}, the three classic usage scenarios are enhanced to immersive communication, massive communication and hyper reliable and low-latency communication (HRLLC). {Beyond these three, there are a group of novel usage scenarios, namely, ubiquitous connectivity, integrated sensing and communication (ISAC) as well as artificial intelligence (AI) and communication.}

It is fairly straightforward to understand the three communication based scenarios, which were evolved from the three pillar scenarios of 5G.
Immersive communication covers a wide range of novel use cases, including extended reality (XR) and  holographic communications, which would {mainly} require larger transmission bandwidth than eMBB in 5G.  
%They include a variety of possible deployment scenarios such as hotspots, rural and urban. 
HRLLC extends the boundary of URLLC by covering more specialized use cases that are expected to have more stringent requirements mainly on reliability and latency, where failure in meeting these requirements could lead to severe consequences for {envisioned} applications. Some typical use cases include remote tele-surgery, fully autonomous driving as well as industrial control and operation.
Massive communication, the evolution of mMTC, intends to support connectivity of a huge number of devices, possibly in a relatively small area. Emerging use cases belonging to this scenario include dense sensor networks for industries, metaverse, logistics and transportation. %{Note that a particular use case may reside under multiple usage scenarios when it needs to be supported by a full range of capabilities.}

Ubiquitous connectivity, as a new usage scenario, is intended to enhance connectivity to bridge the digital divide. {The} focus of this usage scenario is to address presently uncovered or scarcely covered areas, particularly rural, remote and sparsely populated areas, through possible interworking with other access systems. Typical use cases include {Internet} of things and mobile broadband communication.

The usage scenario of ISAC offers new possibilities for end users in terms of what they can expect from a network. Sensing as a network service will become a reality, providing more options to the market as well as boosting relevant vertical industries. Use cases such as imaging of the surrounding environment, mapping, gesture and activity recognition, target detection and tracking, security surveillance and navigation, and even social welfare such as disaster monitoring can all {potentially} be empowered by 6G. This makes ISAC an attractive usage scenario.
The major challenge in the way for ISAC to become a successful usage scenario is how to balance communication and sensing services as they compete for the same set of resources. Historic knowledge and experience will not help architects and designers to understand this trade-off, since previous generations do not consider sensing, nor is sensing provided as a service.  It is also worth discussing how ISAC will eventually become a successful business model that could be profitable to network operators so that they will be motivated to {offer it}, given that there are always other alternatives such as radar systems.

AI can also be integrated with {communications} and provide a portfolio of novel services to both private users and vertical industries. AI-as-a-service is now quite common for many industries, providing  options to deploy in-house AI using off-the-shelf tools. In this way, vertical industries can enjoy the advantages of AI without hiring engineers and without purchasing their own hardware or collecting data for training the models. Taking one step ahead, 6G will offer off-the-shelf services integrating both AI and communication functions, supporting vertical industries to fulfill their jobs in a more efficient way. One example would be security surveillance, where a camera records live video of a given scene. Instead of uploading everything to the server to store, an AI service can first process the live video to identify if there is anything abnormal. {If so}, AI can identify the objects which are suspicious, and then upload the recorded clip to the server for human inspection. By combining AI and communication, such a service is provided directly by one supplier, specifically the network operator, saving customers time and efforts to purchase two separate services and linking them together. 

\subsection{Overarching Aspects}
In Fig. \ref{fig:usage}, there are four overarching aspects in the context of all usage scenarios. 
%These four aspects are not considered as usage scenarios, but are treated as critical factors need to be considered in 6G. 
In particular, sustainability reflects the need for greener communication networks, not only by reducing the power consumption but also through manufacturing equipment and devices with longer life cycle and less environmental impact. There is also broad consensus that security and privacy are vital, since significantly more personal and industrial data will be transmitted through 6G networks. When the communication society rushes from 5G to 6G, it is also worth noting that many underdeveloped regions in the world still rely on very old networks. Thus, it remains the mission of the IMT industry to ensure digital inclusion for all. As intelligence is needed in every part of the network, from servers and base stations (BSs) to user equipment (UE) and sensors, intelligence will become ubiquitous across all usage scenarios. These four aspects {need to} be taken into account when designing 6G systems.

\subsection{Capabilities}
{At the time of writing, the capabilities of 6G are only specified by ranges of target values}. More detailed metrics are expected to be decided during the TPR phase of ITU-R WP5D.
Nonetheless, these capabilities can serve as a rough guideline for future system designers. 

\begin{figure}[t]
\centering
\includegraphics[width=0.85\linewidth]{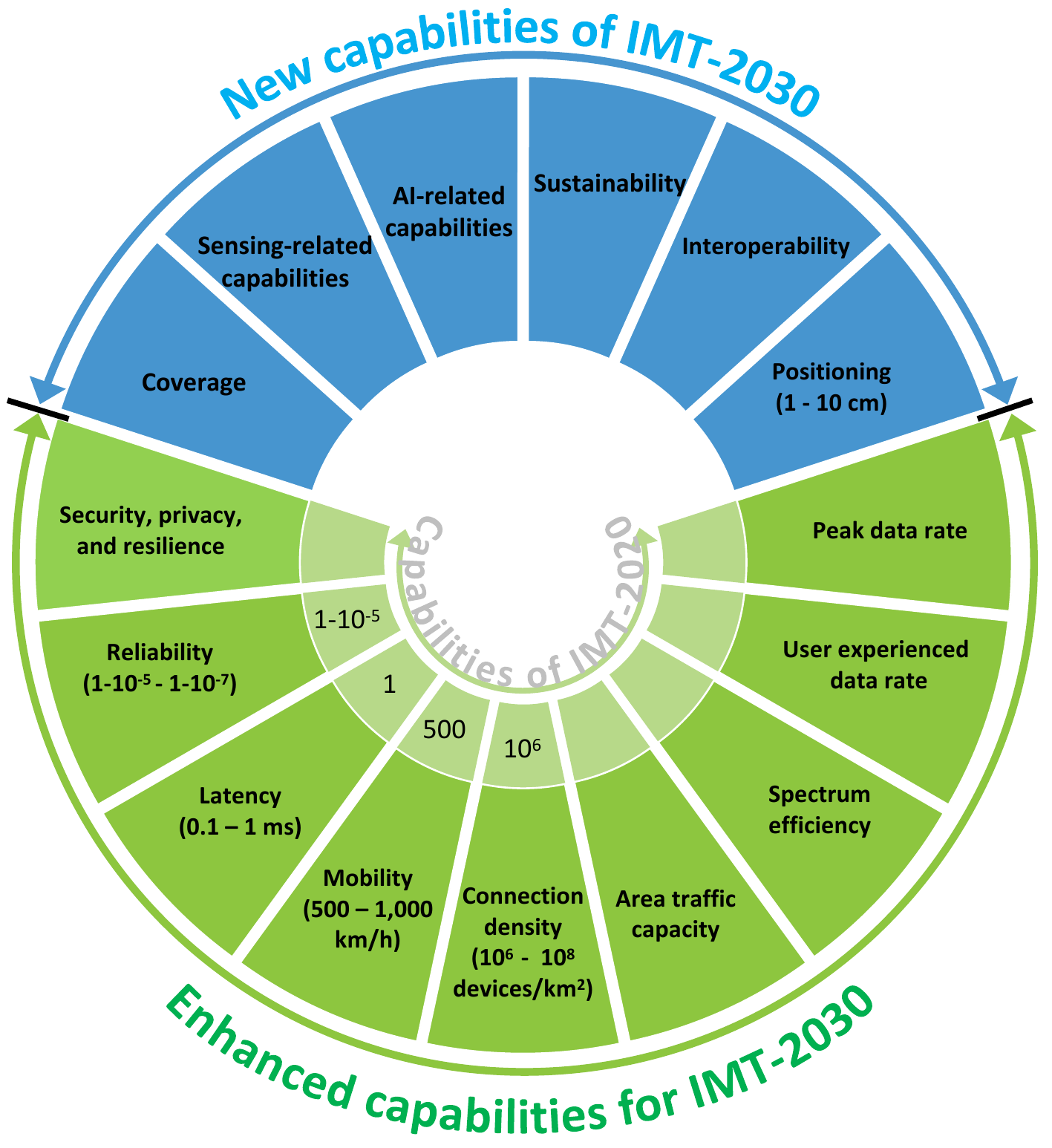}
\caption{An illustration of 6G capabilities compared to 5G. The range of values given for capabilities are estimated targets for research and investigation of 6G. (Source: \cite{ITU_framework})}
\label{fig:capabilities}
\end{figure}

All the capability items endorsed by ITU-R as well as their range of values are presented in Fig. \ref{fig:capabilities}. 
%Note that 6G will have six new capability items, which are depicted in blue color. 
As some of the capabilities are complicated to be quantified, measured and evaluated, they are {qualitatively stated} \cite{10298069}.
It can be seen that 6G is expected to provide substantially enhanced capabilities in almost all dimensions. It is worth noting that the ITU-R will treat all possible values within the range with equal priority, including upper and lower bounds. {Note that some metrics do not present a clear range of values in Fig. \ref{fig:capabilities} due to no consensus in  WP5D, while some examples are given in the text of the recommendation} \cite{ITU_framework}.
The most enhanced capabilities include peak data rates (200 Gbps is given as an example compared to 20 Gbps in 5G), user experienced data rates (potentially more than five times greater than 5G), area traffic capacity (potentially more than five times greater than 5G), reliability and connection density. This stronger capability portfolio will help boost the user experience to the next level, and empower vertical industries as well.

\begin{table}[t]
    \caption{Representative capability items corresponding to the six usage scenarios}
    \label{Tab:cap_correspond_to_usage}
    \centering
    \begin{tabular}{|p{3.5cm}|p{3.5cm}|}
    \hline
        Usage scenarios & Representative capabilities  \\ \hline
        Immersive Communication & Peak data rate, user  experienced data rate, area traffic capacity, spectrum  efficiency, latency, mobility  \\ \hline
        Hyper Reliable and Low-Latency Communication & Latency, reliability \\ \hline
        Massive Communication & Connection density, area traffic capacity, energy efficiency \\ \hline
        AI and Communication  & Applicable AI-related capabilities, security and resilience  \\ \hline
        Integrated Sensing and Communication & Positioning accuracy, sensing-related capabilities  \\ \hline
        Ubiquitous Connectivity &  Coverage \\ \hline
    \end{tabular}
\end{table}

Capability items are usually linked to usage scenarios during the TPR and evaluation studies, when the ITU-R will start to study evaluation methodologies and test cases to verify the capabilities. Table \ref{Tab:cap_correspond_to_usage} presents the correspondence between the six usage scenarios and capability items in our view. As the concept of usage scenarios is quite broad, a large set of capability items can be found relevant to them, while only some representative ones will be tested and verified. 

\section{Technical Enablers}\label{sec:FTT}
\begin{figure*}[t]
\centering
\includegraphics[width=0.98\linewidth]{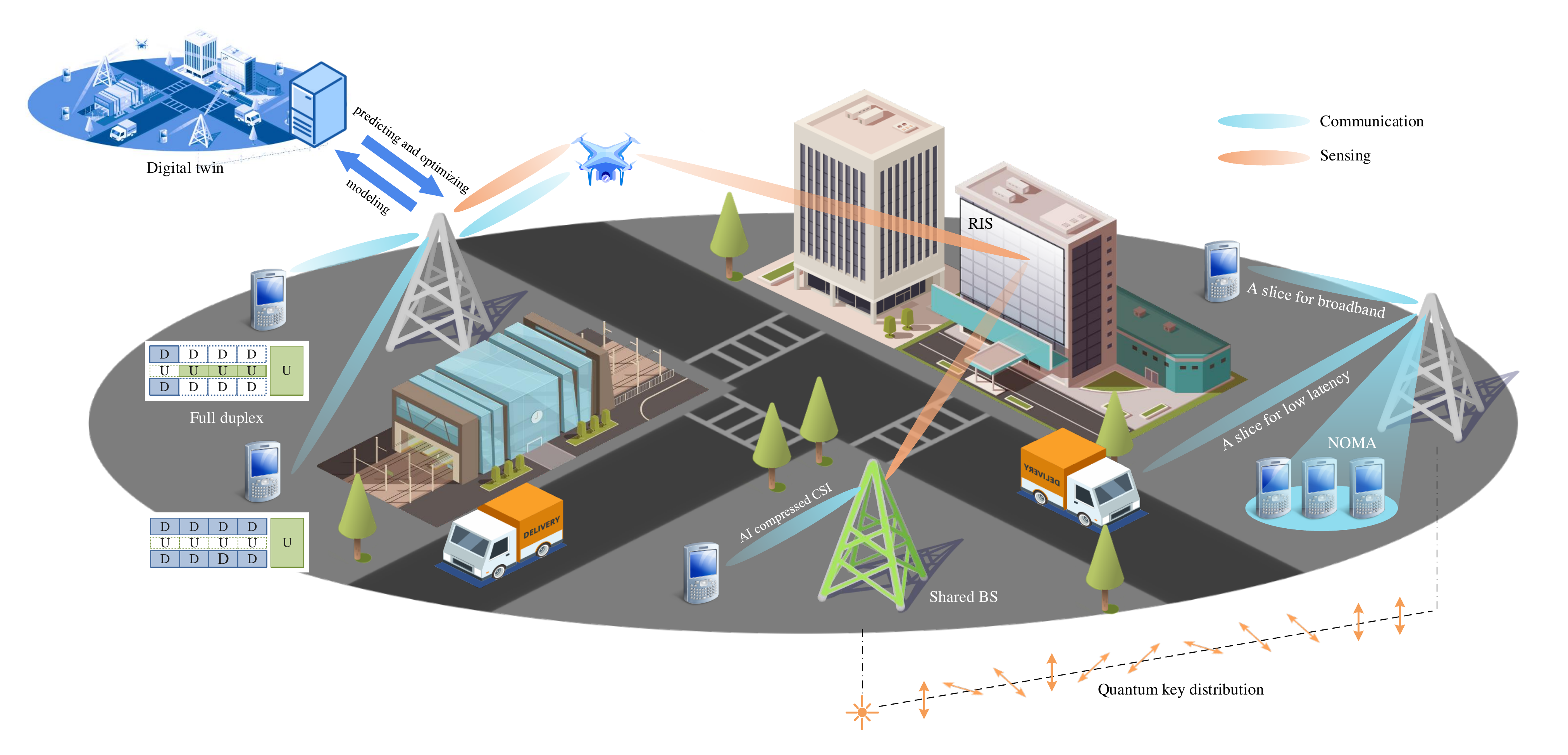}
\caption{Prospective 6G technology enablers.}
\label{fig:enablers}
\end{figure*}

To best serve the six usage scenarios, 6G needs novel architectures, technologies and designs. In general, many existing and novel technologies will play important roles in 6G and contribute to certain aspects as depicted in Fig. \ref{fig:enablers}, and summarized in Table \ref{Tab:techs} in terms of  benefits and challenges. Among them, some of the most promising candidate technologies are analyzed next.

\begin{table*}[t]
    \caption{{Technologies to impact 6G}}
    \label{Tab:techs}
    \centering
    \begin{tabular}{|p{2.4cm}|p{7cm}|p{7cm}|}
    \hline
        Technologies & Benefits  & Challenges to implement  \\ \hline
        AI & Advanced capabilities in signal processing and inference. & High energy consumption and a harmonized architecture to integrate AI into networks. \\ \hline
        ISAC & Abilities to sense the environment re-using communication infrastructures.  & True integration of sensing and communication on hardware, algorithm and signal level. \\ \hline
        RIS & Improving communication performances as well as enabling novel services in a cost-effective and energy-efficient way. & Potential out-of-band reradiation and interference.  \\ \hline
        Full duplex & Potentially doubling system capacity and spectral efficiency. & Demanding high antenna isolation and advanced signal processing techniques. \\ \hline
        RAN slicing & Providing tailored services to users of diversified needs. & Managing slices in a real-time manner and ensuring quality of service for all slices. \\ \hline
        RAN infrastructure sharing & Greatly saving costs and investment for operators. & Managing the infrastructures in a fair and effective way. \\ 
        \hline
        Non-orthogonal multiple access (NOMA) & Significantly improving the spectral efficiency. & Interference cancellation. \\  \hline
        Digital twin & Abilities of predicting and managing the network behavior in a real-time manner. & High complexity and cost to model the whole system as well as the users and the environment. \\  \hline
    \end{tabular}
\end{table*}

\subsection{AI}
AI has revolutionized many industries and will create significant impact in wireless communications as well. The interplay of AI and 6G will be in two ways: AI for communication and communication for AI, both of which will open new possibilities for 6G networks.

The communication systems have become larger and more complex every generation and many tasks would benefit from introducing native AI into them, such as channel estimation. %As the communication frequency becomes higher, more antenna elements are needed to support large scale multiple-input multiple-output (MIMO) systems 
To enable multi-user transmissions in the face of increased connection density, large scale multiple-input multiple-output (MIMO) systems supporting a high number of data streams are needed. This in turn leads to increased pilot overhead, as well as potentially escalating channel estimation and feedback complexity. Conventional methods of feeding back the channel state information from the UEs to the BS usually lead to design dilemmas in terms of carefully balancing the accuracy and overhead, when the number of antenna elements increases. This challenging situation can be solved by {a two-sided AI model deployed on both the UE and the BS. Specifically, the UE employs an AI-based encoder to generate compressed feedback information, while a corresponding AI-based decoder at the BS is used to reconstruct the CSI}. Other examples of AI boosting the network performance and reducing costs include sensing, resource allocation, scheduling as well as network planning and optimization \cite{8957702}.

On the other hand, AI also needs assistance from the network, in support of novel usage scenarios. Enhanced uplink transmission is needed for accommodating the large volume of data required for training AI models, changing the previous downlink-heavy communication paradigm. The core network also has to evolve, as more computational tasks will be performed at the edge instead of by a central server, calling for new designs concerning both the architecture and transmission protocols. For instance, a service based architecture and accurate quality-of-service (QoS) control can be considered, treating AI as a service to other network features and improve their performance metrics \cite{10210204}. As there exists a variety of AI models and services, each service may require a certain type of communication and computing resources, which can benefit from RAN slicing.

From an architectural point of view, communication systems are designed based on a layered structure for the sake of reducing the design complexity. {Facilitated by the powerful capabilities of AI, cross-layer designs that can achieve a global optimum deserve further research.}

%\subsection{Integrated Sensing and Communication (ISAC)}
\subsection{ISAC}
Similar to AI, ISAC has been identified as an independent usage scenario, which has to be supported by strong technical enablers such as physical layer ISAC techniques. 
To some extent, 5G networks nowadays can also serve as sensors, by utilizing some reference signals in the preamble \cite{liu2023ISAC}, but the true power of ISAC will only be unleashed when the fundamental system architecture takes sensing into account. 
To achieve this goal, 6G networks have to intrinsically amalgamate these two systems into a single one, minimizing the overall cost, size and power consumption, while meeting the target requirements of both functions. This integration will happen not only at the physical layer, but also at higher layers and will potentially impact the core network as well, since new nodes might be needed to coordinate sensing and communication functions.

On a theoretical level, there are challenges which have to be overcome, including the associated electromagnetic modeling (especially for the near-field), optimal waveform design, joint beamforming, power saving mechanisms, timing and synchronization as well as flexible duplexing and control. How to balance the need of high accuracy sensing and robust communication is another key issue to be studied. In this context, it would be beneficial to develop a unified channel model suitable for both purposes and to conceive cross-domain waveform designs  \cite{10556621}.

Indeed, significant standardization efforts are needed to specify new air interfaces for the integrated system. How to modify the core network is another challenge, especially when backward compatibility with legacy systems is considered. Depending on how the communications and sensing functions are integrated, different levels of mutual enhancement or cross link interference (CLI) can be studied. The frequency bands for sensing needs to be selected carefully, considering potential inter-operator interference if two operators own adjacent bands.

\subsection{Reconfigurable Intelligent Surface (RIS)}
RIS is a new technology that can dynamically manipulate electromagnetic waves between transmitters and receivers, converting the wireless environment into a service \cite{9475160}. %An RIS is usually a flat surface constructed of a variety of passive scattering components, each of which may independently impose a certain phase shift and sometimes an amplitude gain on the input electromagnetic waves. 
The propagation of the reradiated electromagnetic waves can be modified by carefully changing the phase shifts (and amplitudes) of the re-configurable elements of RIS. 

Compared to relays and repeaters used in 5G, RISs can be constructed utilizing {nearly} passive components rather than expensive active ones like power amplifiers, leading to low implementation costs and energy use. Consequently, RIS is considered as a sustainable and ecologically friendly technological solution \cite{WCM_RIS_Standards}. In addition, thanks to the large aperture size and large number of reflective elements, RIS can form sharp directional beams and provide better resolution for positioning and sensing services, as well as extra degrees of freedom in the near-field. Passive RISs face challenges including the inability to carry out channel estimation, signal regeneration, and amplification as a result of the lack of power amplifiers and digital signal processing capabilities. Novel types of RIS, such as active RIS that employs active elements to facilitate channel estimation or power amplification, are viewed as a promising solution, along with advanced signal processing techniques, compressed sensing and AI. Another relevant topic is channel estimation when the coherence time is short, which can be tackled by having RISs directly controlled by the BS and proper network planning and optimization.
%These eventually lead to an intrinsic trade-off between the coverage range, size, and the number of RIS elements that must be deployed on it. 

\subsection{Full Duplex Operation}
Spectrum efficiency has become an ever critical capability of communication systems since more services are competing for bandwidth which is a scarce resource. There are some sophisticated techniques to enhance duplex flexibility in 5G, such as the dynamic time division duplexing. However, 6G would demand further enhancements.
Full duplex allows downlink and uplink transmissions to happen simultaneously using the same frequency bands. Being able to double the system capacity theoretically makes full duplex a competitive candidate technology for 6G \cite{9110914}.  However, self-interference is a critical challenge for full duplex operations, because the high-power transmit signal might completely overwhelm the low-power receive signal.

To elaborate, when the transceiver transmits, its own receive antennas will also receive the transmitted signals, which are much stronger than the desired signals due to the difference in the transmission distances. A potential solution to self-interference mitigation is successive interference cancellation (SIC), in both the analogue and digital domain. SIC still needs further enhancement when dealing with massive MIMO systems, which will be common for 6G. From a hardware perspective, high isolation antennas are one of the directions to help alleviate self-interference. {Subband full duplex (SBFD), which utilizes full duplex only at the BS, is seen as a crucial evolution step forward. In SBFD, different types of interference, including BS self-interference, BS inter-subband CLI and UE-UE inter-subband CLI, can be mitigated by physical antenna isolation and digital signal processing. These interference mitigation methods have been proven to be valid in practical environments by trials \cite{10273621}.}

\subsection{RAN Slicing}
Network slicing is a novel design principle of the network architecture that allows operators to create several logically independent networks utilizing a shared physical infrastructure \cite{9314166}. These logical networks are called slices, which can be configured for different users having different requirements. The slices differ from each other not only by the time and frequency resources, but also by the quality-of-service they support. 
%Deployed on the same physical infrastructure, one slice can serve HRLLC with very low latency while the other supports immersive communication with ultra-large bandwidth.

RAN slicing will help vertical industries to deploy their own network by ordering a slice from the operator at a fair price. {For instance, video based quality inspectors in smart factories would demand a slice with large bandwidth in the uplink, while robots in the factory would benefit from a slice with ultra low latency for real-time controlling.}
To support totally different requirements using different slices, 6G has to be designed from the beginning in a flexible way that is easy to configure and manage. 
The key challenge is how to dynamically configure different RAN slices and how to manage them effectively in a timely manner. Again, AI empowers potential opportunities to solve this issue.

\subsection{RAN Infrastructure Sharing}
The sharing of RAN infrastructure by multiple operators can help them build 6G networks at a reduced cost as well as create new business models for them. 
In 5G, RAN infrastructure sharing has already been verified when two major Chinese operators decided to build and share their network infrastructure.
However, RAN sharing faces challenges in terms of transparency, reliability, protection of user privacy and efficient maintenance.
A fair and transparent way for multiple operators to be able to track the cost and usage of the network infrastructure is a stepping stone to make such sharing possible. A promising solution for this problem is blockchain, which can establish a decentralized ledger that is extremely hard to manipulate and thus, trusted by all parties. ITU established a focus group on distributed ledger technologies with 5 reports and 3 technical specifications published \cite{ITU_digital_ledger}. These works on the concepts, ecosystem, use cases, reference architecture, assessment criteria, standardization landscape and regulatory framework of distributed ledgers will be valuable to future RAN infrastructure sharing.

Despite the challenges ahead, RAN sharing will become a trend for 6G and contribute to provide affordable connectivity to more regions and users worldwide, helping to overcome the digital divide.

\section{Spectrum for 6G}
The new generation of IMT will be built upon advanced technologies, but even more fundamentally, it will rely on beneficial spectral band selection.
It is anticipated that some key usage scenarios and use cases of 6G would require high bandwidths, say for instance, about 0.75 GHz to support high-accuracy ISAC, 1 GHz to facilitate wide-area XR and 1.1 GHz for high-resolution holographic communications. This essentially means that to fully unleash the power of 6G, 500 to 750 MHz new spectrum is needed, in addition to ‘re-farming’ the existing spectrum of previous IMT systems.
To enable all the usage scenarios and support the above novel techniques, additional spectrum has to be assigned for IMT systems during the world radio conference 2023 (WRC-23) and WRC-27.

Following the WRC-23, the spectrum allocations for IMT services in this 4-year study cycle have become clear \cite{WRC23}. The 3.5 GHz band (3.3 - 3.8 GHz) has been ratified across Europe, the Middle East and Africa, marking a significant milestone in the advancement of IMT systems.
The 6 GHz band (6.425 - 7.125 GHz), previously identified as unlicensed, has been allocated to licensed IMT globally. Technical conditions for 6 GHz spectrum are now being globally harmonized for paving the way to expanded mobile capacities. 
The 10 - 10.5 GHz band is the third new band assigned to IMT both in North- and South-America, while certain restrictions apply to protect services using adjacent bands, including the Earth-exploration satellite service. This band is only intended for micro-cell {BSs}, and both the maximum equivalent isotropically radiated power as well as the total radiated power is limited.
It is worth noting that the WRC-23 allocates spectrum to IMT services without specifying a wireless generation, so it is plausible for some countries to start using the aforementioned bands for 5G-Advanced. Besides the newly allocated bands, the bands 470 - 694 MHz and 4800 - 4990 MHz that are already identified for IMT in some regions are now adopted in more regions globally. Note that all IMT bands are treated as licensed bands.

The agenda of WRC-27 was also defined during WRC-23, including the consideration of bands supporting future networks and their features. As per the final act of WRC-23 \cite{WRC23}, the following bands will be studied in terms of whether or not to be further allocated to IMT at WRC-27: 4400 - 4800 MHz, 7125 - 8400 MHz and 14.8 - 15.35 GHz. Potential new bands might be allocated for IMT from these three ranges, and no other frequencies will be considered at WRC-27.

Spectrum allocation is never an issue for a single country or industry. It is a pervasive global topic that involves many stakeholders. Harmonization of the 6G spectrum and standards is critical for 6G to become successful.

\section{Conclusion}
6G aims {at} providing connectivity for the {whole human society} and overcoming the digital divide while facing diverse demands as well as supporting enhanced capabilities than any previous generation.
This goal will be achieved by novel technology enablers, new spectrum allocations and a continuous effort to develop a global standard by all stakeholders.

\bibliographystyle{IEEEtran}
\bibliography{Reference}

\begin{IEEEbiography}
[{\includegraphics[width=1in,height=1.25in,clip,keepaspectratio]{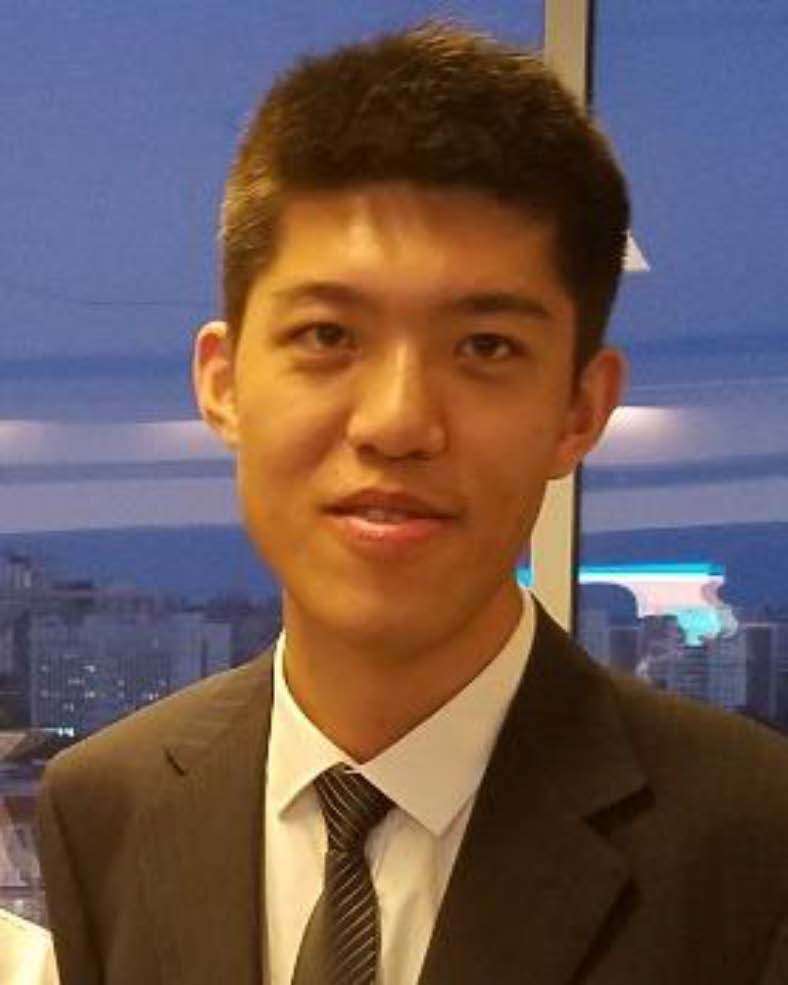}}]{Ruiqi (Richie) Liu} is a master researcher in the wireless and computing research institute of ZTE Corporation. He was deeply involved in specifying 5G standards through 3GPP, where he served as a rapporteur. He is a Voting Member of the IEEE ComSoc Industry Communities Board.
\end{IEEEbiography}

\begin{IEEEbiography}
[{\includegraphics[width=1in,height=1.25in,clip,keepaspectratio]{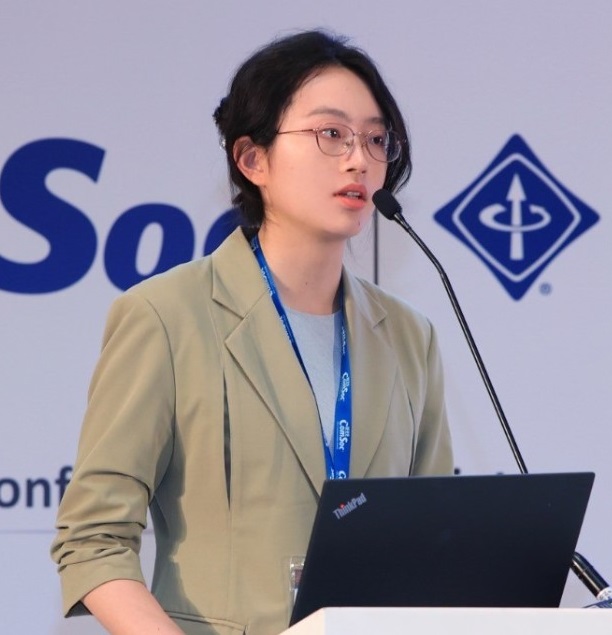}}]{Leyi Zhang} is a researcher with the technology planning department of ZTE Corporation. She actively participates in global standardization activities through ITU, 3GPP, ETSI and CCSA and is a frequent speaker at international conferences.
\end{IEEEbiography}

\begin{IEEEbiography}
[{\includegraphics[width=1in,height=1.25in,clip,keepaspectratio]{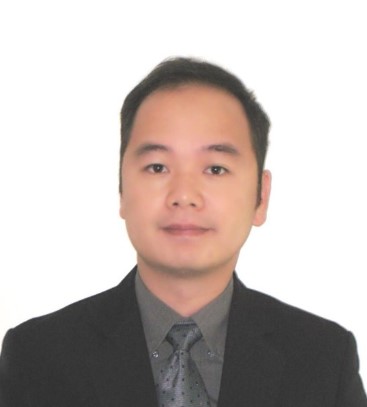}}]{Ruyue Yu-Ngok Li}
is a ZTE fellow and the VP of Radio System Research and Standardization at ZTE Corporation in charge of research and standardization on radio system related 5G/6G technologies. He received his Ph.D. and B.Eng. degrees from the University of Hong Kong and his M.S. degree from Stanford University. He holds over 100 granted patents. %Since he joined ZTE in 2009, he has been actively involved in wireless research and standardization activities including 3GPP RAN standardization. Prior to ZTE, he worked for several telecommunication and semiconductor companies including Qualcomm and Marvell Semiconductor on projects related to 2G/3G baseband algorithm design and LTE standardization. His recent research interests include massive MIMO, mmWave communications, beam management, reconfigurable intelligent surfaces, AI/ML applications on MIMO, URLLC, green communication and interference coordination.
\end{IEEEbiography}

\begin{IEEEbiography}
[{\includegraphics[width=1in,height=1.25in,clip,keepaspectratio]{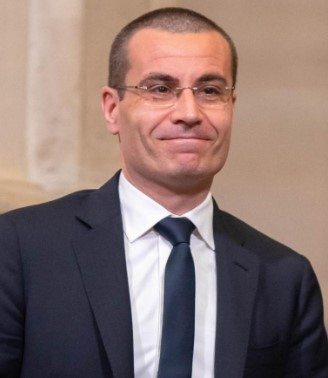}}]{Marco Di Renzo}
	  is a CNRS Research Director and the Head of the Intelligent Physical Communications group in the Laboratory of Signals and Systems at CentraleSupelec - Paris-Saclay University. He is a Fellow of the IEEE, and he serves as the Director of Journals of the IEEE Communications Society.
\end{IEEEbiography}

\end{document}